# Unconventional Ferroelectricity in Violation with Neumann's Principle

Junyi Ji[1,2], Guoliang Yu[1,2], Changsong Xu[1,2]*, and H. J. Xiang[1,2,3]*

[1]Key Laboratory of Computational Physical Sciences (Ministry of Education), Institute of Computational Physical Sciences, and Department of Physics, Fudan University, Shanghai 200433, China

[2]Shanghai Qi Zhi Institute, Shanghai 200030, China

[3]Collaborative Innovation Center of Advanced Microstructures, Nanjing 210093, China

[†]J.J. and G.Y. contributed equally to this work.

Email: csxu@fudan.edu.cn, hxiang@fudan.edu.cn

**Abstract**

   The physical properties of crystals are governed by their symmetry according to Neumann's principle. However, we present a case that contradicts this principle wherein the polarization is not invariant under its symmetry. We term this phenomenon as unconventional ferroelectricity in violation of Neumann's principle (UFVNP). Our group theory analysis reveals that 33 symmorphic space groups have the potential for UFVNP, with 26 of these symmorphic space groups belonging to non-polar groups. Notably, the polarization component in UFVNP materials is quantized. Our theory can explain the experimentally proven in-plane polarization of the monolayer α-$In_2Se_3$, which has $C_{3v}$ symmetry. Additionally, we employ first-principles calculations to demonstrate the existence of UFVNP in $T_d$ phase AgBr, which was not initially anticipated to exhibit polarization. Thus, UFVNP plays an integral role in characterizing and exploring the possible applications of ferroelectrics, significantly expanding the range of available materials for study.

**Main text**

   Ferroelectricity is a property exhibited by insulators, characterized by a spontaneous electric polarization that can be reversed by an applied electric field. This property finds a wide range of applications, such as piezoelectric devices [1-3], non-volatile memory [4-6], and photovoltaics [7-12]. In the field of ferroelectrics, Neumann's principle, which states that the physical properties of a crystal must be invariant under the same symmetry operations that leave the crystal structure unchanged [13], is widely applied. Perovskite ferroelectrics such as rhombohedral phase $BaTiO_3$ [14-17] exhibit polarization that aligns with the three-fold rotational axis. Two-dimensional (2D) MX (M=Ge, Sn; X=S, Se, Te) [18-23] with $Pmn2_1$ space group exhibits an in-plane polarization along the two-fold screw axis. Almost all ferroelectric materials' polarization directions follow Neumann's principle. Previously, ferroelectricity was thought to exist only in 10 polar point groups: $C_1$, $C_2$, $C_s$, $C_{2v}$, $C_4$, $C_{4v}$, $C_3$, $C_{3v}$, $C_6$, $C_{6v}$.

In 2017, monolayer α-$In_2Se_3$ was predicted, by Ding *et al.*, to have both out-of-plane and in-plane polarization [24]. Later, ferroelectric one-layer α-$In_2Se_3$ was prepared and polarization was reversed by moving the central Se layer [25-28]. The in-plane polarization of monolayer α-$In_2Se_3$ was experimentally observed [27-29]. We noticed that monolayer ferroelectric phase α-$In_2Se_3$ has $C_{3v}$ point group symmetry and should have only out-of-plane ferroelectric polarization according to Neumann's principle. This discovery challenges our understanding of ferroelectricity and its relation with symmetries. So far, why the ferroelectric polarization is in violation of Neumann's principle and how this happens have not been noticed and explored.

In this Letter, we present a theory of UFVNP to reconcile the contradiction between ferroelectric polarization in UFVNP systems and its symmetry. Through group theory analysis, we categorize UFVNP into two types based on whether the systems belong to the polar groups or not. By exhausting all 230 space groups, we show that UFVNP can occur in 33 symmorphic space groups, of which 26 are non-polar. Our findings can be used to swiftly determine whether a material is capable of exhibiting UFVNP. Furthermore, we anticipate the occurrence of a second type of UFVNP in the F-43m phase of AgBr. To support our prediction, we performed first-principles calculations. UFVNP has opened up new avenues for exploring the potential of many non-polar materials as candidates for novel ferroelectric devices.

***Group theory analysis of UFVNP***. Let us illustrate the basic idea of UFVNP with a simple example. Consider a 2D material with a triple layer structure ($F_1$-M-$F_2$ from bottom to top) depicted in Fig. 1(a-c). $F_{1,2}$ and M denote the fixed and movable atoms (layers), respectively. The movable atom M lies at the center between $F_1$ and $F_2$ of the structure in the high symmetry phase H (space group P6/mmm, point group $D_{6h}$) (see Fig. 1(b)). Since there is only one M atom in the unit cell, all symmetry operations in H will maintain the position of the M atom. The low symmetry phases $L_1$ and $L_2$ (space group P-6m2, point group $D_{3h}$) can be seen as translating the movable atom M to the center of the triangle of $F_{1,2}$ atoms with an in-plane displacement, which is shown by the blue and red arrows in Fig. 1(b). In the phase transition $L_1$-H-$L_2$, the polarization variation, $\Delta P$, has an in-plane component owing to the non-integer in-plane displacement, as shown by the green arrows in Fig. 1(b). According to Neumann's principle, however, the polarization of the two low symmetry phases, $P_1$ and $P_2$, must be zero since the space group of the two low symmetry phases is P-6m2 (point group $D_{6h}$). As a result, the in-plane polarization originating from atomic displacement contradicts the group of the low symmetry phase. We name this phenomenon as unconventional ferroelectricity in violation of Neumann's principle (UFVNP).

Let us discuss the simplest case of UFVNP with only one movable atom in the group theory way. To distinguish the different positions of the movable atom M in the three phases, positions in the high symmetry phase and low symmetry phases are set as $M_0$ and $M_{1,2}$, respectively. Designate the groups of the high symmetry phase and low symmetry phases as $G_H$ and $G_L$, respectively. Since all symmetry operations in $G_H$ will keep $M_0$ invariant, they also keep H-M, the structure consisting of the fixed atoms, invariant. The symmetry operation $g_L \in G_L$ turns both H-M and $M_{1,2}$ into themselves. In other words, $g_L$ is the symmetry operation in $G_H$ that maintain $M_{1,2}$ invariant and $G_L$ is the site-symmetry group of Wyckoff position $M_{1,2}$ in $G_H$. If the atomic displacement $\Delta M$ is not invariant under $G_L$, UFVNP occurs. There are two types of UFVNP: type-I. $G_L$ is a polar group but $\Delta M$ does not point to the direction invariant under $G_L$, such as $In_2Se_3$; type-II. $G_L$ is a non-polar group but $\Delta M$ is nonzero. The second type of UFVNP is even more surprising because it suggests that polarization can arise in systems previously thought to be nonpolar. Moreover, the mechanism is completely different from past studies[30]. Materials that were previously ruled out as candidates for ferroelectric materials due to their symmetry could now have the potential to become ferroelectric.

It is worth noting that due to the periodicity of the lattice, there are many equivalent $\Delta M$ as shown by the green arrows in Fig. 1(b). These arrows show three possible $L_1$-$L_2$ phase transition paths, which indicates the high symmetry phase H may not exist in the phase transition. Since $G_H$ can be interpreted as the group of H-M, $M_0$ can always be set as Wyckoff position (0,0,0) (even if there is an atom in this position) to construct the non-existent H, which is designed to facilitate the reader's understanding of the phase transition and the relation between low symmetry phases. The key group theory elements in the UFVNP model are the groups $G_H$, $G_L$, and $M_{1,2,...}$, corresponding to the fixed structure H-M, the ferroelectric phase L and the positions of the movable atoms in L.

Next, we will find all the possible space groups that may exhibit the simplest UFVNP. If the system has a non-symmorphic operation, such as the gliding mirror, the operation on a movable atom will generate another movable atom. This is contradicted by the fact that only one of the atoms is movable. Therefore, only symmorphic space groups need to be considered. Since the energies of low symmetry phases are equal, $M_{1,2,...}$ are those Wyckoff positions that have different coordinates but share the same Wyckoff letter. The space groups exhibiting UFVNP are found by the following steps: 1) list the Wyckoff positions of a symmorphic space group ($G_H$); 2) select a Wyckoff letter that has at least two different Wyckoff positions ($M_{1,2,...}$) and figure out the site-symmetry group ($G_L$) of each Wyckoff position; 3) judge whether this vector $\Delta M$ is invariant under $G_L$, if not, UFVNP may occur in the system with space group $G_L$, otherwise only the conventional ferroelectricity can exist. By exhausting the Wyckoff positions and the corresponding site-symmetry group in 73 symmorphic space groups [31], we find all possible space groups for UFVNP, as listed in Tab. 1 and Tab. S1. There are 52 possible

$G_H$ and 33 possible $G_L$, where 26 of $G_L$ are non-polar and correspond to the type-II UFVNP.

The modern theory of electric polarization shows that electric polarization is a lattice rather than a vector. Based on this theory, we proved that the polarization component of UFVNP is quantized [see SM]. For systems with three-fold rotation, the quantized in-plane polarization components can be (m/3,n/3), where m and n can be 1 or 2. This is consistent with a previous study [32]. The quantized polarization components can be half-integer with inversion, two-fold rotation and mirror.

Finally, we'll go over how to use the tables to tell if a given structure has UFVNP and how to find the other symmetry related low-symmetry phase(s) (see Fig. 1(d)). We assume that the low symmetry phase $L_1$ is given (e.g., in a crystal dataset). Then we carry out the following procedure: 1) Check if its group is one of the $G_L$ in Tab.1. If yes, 2) Find the corresponding $G_H$ in Tab. S1. 3) Remove an atom with site-symmetry group $G_L$. 4) Check if the new structure belongs to $G_H$. If no, go back to 3), else 5) Apply the symmetry operation in $G_H/G_L$ to $L_1$ to obtain the other low symmetry phase(s), $L_{2,...}$. It should be emphasized that the Wyckoff position of an atom in the same crystal may be different with different unit cell settings. Thus, it is not recommended to find the other low symmetry phases using the Wyckoff positions (Wyckoff letters).

***UFVNP in Monolayer α-In₂Se₃.*** Monolayer α-$In_2Se_3$ is a 2D material with a quintuple layer structure Se-In-Se-In-Se as shown in Fig. 2(a) and its space group is P3m1 (No. 156). During the phase transition shown in Fig. 2(a), the middle Se atom moves from directly above the lower In atom to directly beneath the upper In atom. The in-plane displacement is $\frac{1}{3}a + \frac{1}{3}b$ and corresponds to a non-zero quantized in-plane polarization, which contradicts the P3m1 symmetry of the system. Monolayer α-$In_2Se_3$ has a type-I UFVNP.

Now we apply the UFVNP theory to the monolayer α-$In_2Se_3$. The low symmetry phase $L_1$ belongs to $G_L$=P3m1 and the movable atom M is the middle Se atom. From H-M, which is a four layers structure without the middle Se atom (Se-In-In-Se), one can figure that $G_H$=P-3m1 (No. 164 space group). Apply an operation in $G_H$ but not in

$G_L$, such as inversion with center at the origin, to $L_1(L_2)$, one can obtain the other low symmetry phase $L_2(L_1)$.

DFT calculations are further performed to verify the contradiction between the in-plane polarization of monolayer α-In$_2$Se$_3$ and its symmetry. The minimum energy pathway between $L_1$ and $L_2$ is calculated using the climbing image nudged elastic band (CI-NEB) method [33,34] and the barrier height is 68 meV/f.u (see Fig. 2(b)), which is quite close to the value in Ref. [23]. Next, we focus on the in-plane polarizations of monolayer α-In$_2$Se$_3$. The in-plane polarizations are calculated using the Berry phase approach [35-37]. As shown in Fig. 2(b), the magnitude of the in-plane polarizations of monolayer α-In$_2$Se$_3$ varies continuously along the $L_1$-$L_2$ pathway. Among them, the in-plane polarizations of $L_1$ and $L_2$ phases along the [110] direction are 2.37 and -2.37 $e$Å per unit cell, respectively. The quantum $Q = eR$ along the [110] direction, where $e$ and $R$ are the charge of the electron and lattice along [110] direction, is 7.11 $e$Å. So the polarization is $\frac{1}{3}Q$. However, the out-of-plane component is not quantized. The results are in agreement with previous research [24]. The polarization calculations suggest that the monolayer α-In$_2$Se$_3$ is a ferroelectric with in-plane polarizations, which contradicts the symmetry of P3m1. Such results indicate that the system is Type-I UFVNP and the UFVNP theory is well consistent with DFT calculations.

*UFVNP in AgBr.* Based on the UFVNP theory, a material with non-polar point group can be ferroelectric and this is type-II UFVNP. From Tab. 1 and Tab. S1, it can be seen that the highest symmetry of low symmetry phase $L_{1,2,...}$ possible to have UFVNP is space group 216, F-43m, and the corresponding $G_H$ is space group 225, Fm-3m. A typical F-43m structure is Zinc blende, as shown in Fig. 3(a) with chemical formula AB, A at (0,0,0) (Wyckoff letter 4a) and B at (1/4,1/4,1/4) (Wyckoff letter 4d). Remove B (or A), the left structure H-M belongs to $G_H$=Fm-3m. Apply an operation in $G_H$ but not in $G_L$, such as inversion with center at the origin, to $L_1(L_2)$, one can obtain the other low symmetry phase $L_2(L_1)$. The phase transition of $L_1$-$L_2$ can be seen as atom B moving from (1/4,1/4,1/4) to (3/4,3/4,3/4). When B locates at (1/2,1/2,1/2), the structure coincidentally equals to the high symmetry phase H, a rocksalt structure with space group Fm-3m. It can be seen that in the $L_1$-$L_2$ phase transition, the B atom has a non-integer shifting (1/2,1/2,1/2), corresponding to a quantized polarization. However, the point group of the lower symmetry phase $L_{1,2}$ is the non-polar point group $T_d$. We searched materials with both F-43m and Fm-3m phases in the materials project database [38] and chose AgBr to exhibit the type II UFVNP. It has a typical rocksalt experimental structure with space group Fm-3m [39-41]. Other materials with both F-43m and Fm-3m phases are listed in [SM].

We now discuss the ferroelectricity of AgBr by DFT calculations. By phonon spectra calculation and molecular dynamics simulation, we confirm that the low symmetry phase is dynamically stable and has good thermal stability at room temperature [see Fig. S1]. The lattice constants of the high and low symmetry phases are 5.84 and 6.31 Å, respectively. The low symmetry phase $L_1$ and $L_2$ are two energetically degenerate ground states of the system. The $L_{1/2}$ phase has a lower energy than the H phases, and the energy difference between them is about 74 meV/f.u.

To prove the ferroelectricity of the system, we must show that the system has a switchable spontaneous polarization. Fig. 3(a) shows the kinetic pathways between $L_1$ and $L_2$ phases. The high symmetry phase H locates at an energy local minimum. Therefore, the pathway $L_1$-$L_2$ can be split into two $L_1$-H and H-$L_2$. In the $L_1$-H step, the $L_1$ phase transforms into the H phase by shifting the Br from (1/4,1/4,1/4) to (1/2,1/2,1/2) along [111] direction, which overcomes an energy barrier of 155 meV/f.u. In the H-$L_2$ step, the H phase transforms into another low symmetry phase ($L_2$ phase) by shifting the Br from (1/2,1/2,1/2) to (3/4,3/4,3/4) along the same direction, which only needs to overcome an energy barrier of 76 meV/f.u. The switching process of $L_1$-$L_2$ only needs to overcome the highest barrier, that is, the barrier of 155 meV/f.u. in the first step, which is comparable to that of the traditional bulk ferroelectric material $PbTiO_3$ and $BiFeO_3$ [42]. The energy barrier for the phase transition from $L_2$ to $L_1$ is the same as the above barriers.

Fig. 3(b) shows the evolution of the ferroelectric polarizations along the $L_1$-$L_2$ pathway for the AgBr system. The polarization is calculated using the modern theory of polarization. One can see that the polarization of the system changes continuously as the Br atoms move from (1/4,1/4,1/4) to (3/4,3/4,3/4). The polarization of $L_1$ and $L_2$ phases along the [111] direction are -69.80 and 69.80 $\mu C/cm^2$, respectively. According to the modern theory of polarization, the AgBr systems have a polarization quantum with an amplitude of 69.80 $\mu C/cm^2$ along [111] direction, which is calculated by $eR/\Omega$. Here, $e$, $R$ and $\Omega$ are the charge of the electron, a lattice along [111] direction, and the volume of the unit cell. Therefore, another trend in the ferroelectric polarization of the system is that the $L_1$ and $L_2$ phases have quantized polarization. Overall, these results demonstrate that AgBr exhibits spontaneous polarization and the polarization is switchable, indicating that it is a novel ferroelectric material.

*Summary*. To conclude, we propose a model to describe UFVNP and classify it into two types by whether a UFVNP material's point group is polar or non-polar. In particular, the polarization component in UFVNP is quantized. According to the group theory analysis, UFVNP may exist in 33 symmorphic space groups. We apply the UFVNP theory to explain the ferroelectricity in the monolayer α-$In_2Se_3$. This is a type-I UFVNP and confirms the UFVNP theory. With first-principles calculations, we predicted the type-II UFVNP in AgBr of the F-43m phase. The discovery of UFVNP significantly expands the scope of ferroelectrics and is crucial in exploring their properties and potential applications in different fields.

**Acknowledgment**. This work is supported by NSFC (grants No. 811825403, 11991061, 12188101, 12174060, and 12274082) and the Guangdong Major Project of the Basic and Applied Basic Research (Future functional materials under extreme conditions--2021B0301030005). C.X. also acknowledges support from the open project of the Guangdong provincial key laboratory of magnetoelectric physics and devices (No. 2020B1212060030).

Table. I. Possible symmorphic space groups for ferroelectric phases (low symmetry phases $L_{1,2,...}$) in UFVNP. SG and PG denote the space group and the point group, respectively.

| Polar | | Non-polar | | | | | | | |
|---|---|---|---|---|---|---|---|---|---|
| SG | PG | SG | PG | SG | PG | SG | PG | SG | PG |
| P3 | $C_3$ | P-1 | $C_i$ | I-4 | $S_4$ | R-3 | $C_{3i}$ | R-3m | $D_{3d}$ |
| P3m1 | $C_{3v}$ | P-6 | $C_{3h}$ | P-6m2 | $D_{3h}$ | F23 | T | F-43m | $T_d$ |
| P2<br>C2 | $C_2$ | P2/m<br>C2/m | $C_{2h}$ | P422<br>I422 | $D_4$ | P4/mmm<br>I4/mmm | $D_{4h}$ | P312<br>R32 | $D_3$ |
| Pmm2<br>Cmm2<br>Imm2 | $C_{2v}$ | P222<br>C222<br>F222<br>I222 | $D_2$ | Pmmm<br>Cmmm<br>Immm | $D_{2h}$ | P-42m<br>I-4m2<br>I-42m | $D_{2d}$ | | |

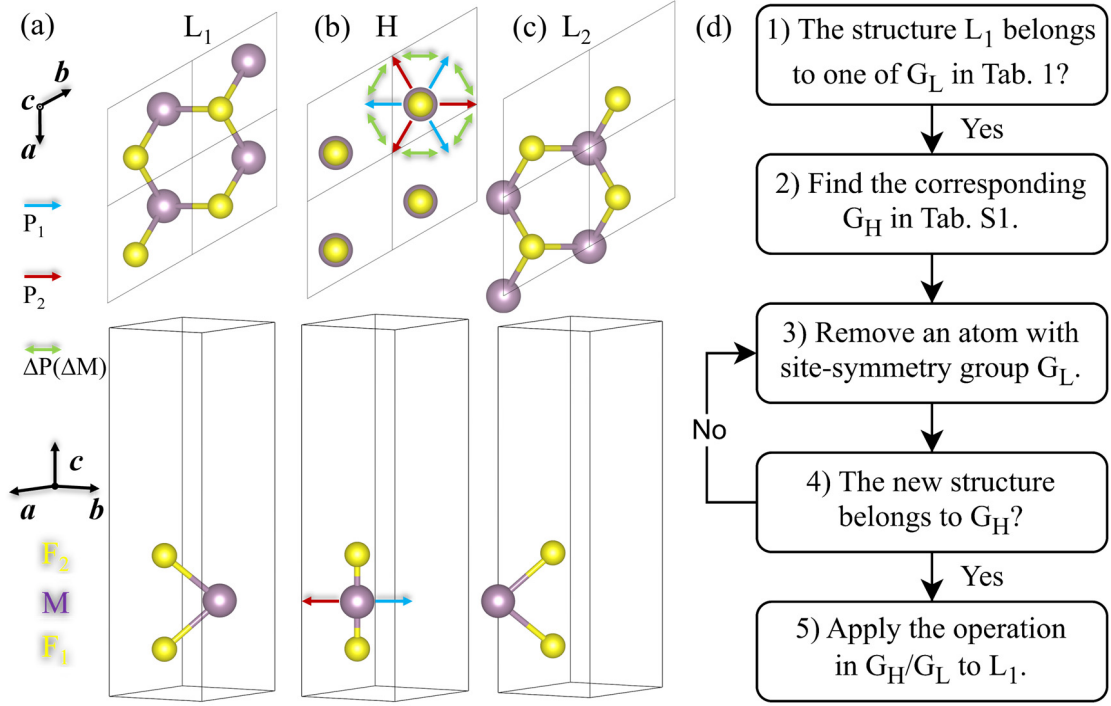

Figure 1. Top and side views of the UFVNP model. (a) low-symmetry phase $L_1$, (b) high-symmetry phase H, (c) low-symmetry phase $L_2$. The black border indicates the unit cell. $F_{1,2}$ are fixed atoms (layers) during the ferroelectric phase transition, and M is a movable atom (layer). The blue and red arrows depict the atomic displacements of M from H to $L_1$ and $L_2$, respectively. The green arrows show ΔM, the atomic displacements of M between $L_1$ and $L_2$. Since only M moves, the blue, red, and green arrows can also represent $P_1$(polarization of $L_1$), $P_2$(polarization of $L_2$), and ΔP(polarization difference between low symmetry phases), respectively. ΔP(ΔM) cannot be invariant under a symmetry operation ($C_{3z}$) of the low-symmetry phase, which leads to the UFVNP. (d) Flow chart for determining whether a material has UFVNP and finding the remaining low symmetry phases (ferroelectric phases).

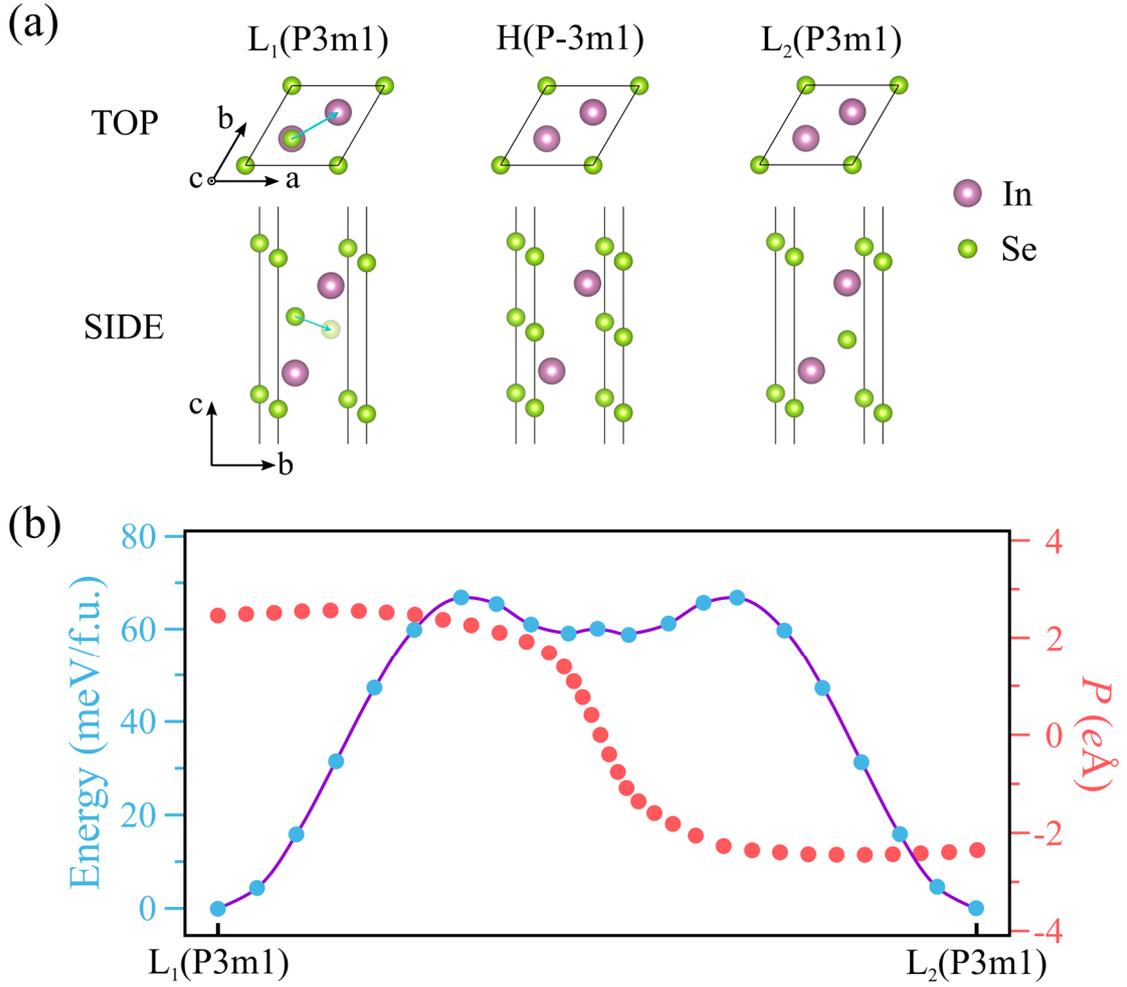

Figure 2. Structure and ferroelectricity of monolayer α-In$_2$Se$_3$. (a) Top and side views of monolayer α-In$_2$Se$_3$, corresponding to the L$_1$, H, and L$_2$ phases, respectively. $\frac{1}{3}\boldsymbol{a} + \frac{1}{3}\boldsymbol{b}$, the in-plane displacement of M (Se) in the ferroelectric phase transition is shown by the blue arrow. (b) NEB calculation of the energy barrier and evolution of the polarization intensity along the path similar to the one in Ref. [24]. The amplitude represents the in-plane polarization magnitude along the [110] direction. The positive and negative signs indicate the polarization toward [110] and the [-1-10] direction, respectively.

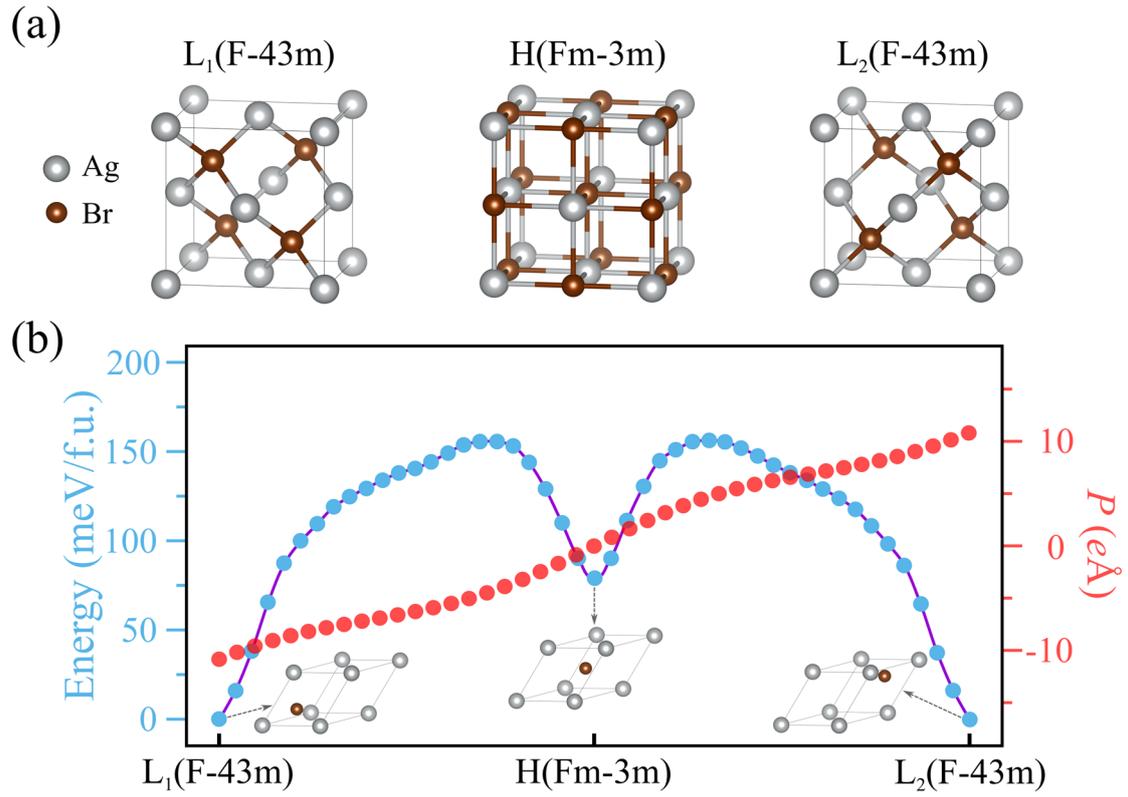

Figure 3. (a) Schematic structure of AgBr during the $L_1$-H-$L_2$ phase transition. The $L_{1,2}$ phase belongs to F-43m and the H phase belongs to Fm-3m. The process can be considered as Br moving along the path $\left(\frac{1}{4},\frac{1}{4},\frac{1}{4}\right) \to \left(\frac{1}{2},\frac{1}{2},\frac{1}{2}\right) \to \left(\frac{3}{4},\frac{3}{4},\frac{3}{4}\right)$. (b) The energy barrier calculated by NEB and the evolution of the polarization along the path in (a), where the unit cells of $L_1$, H, $L_2$ are depicted.